\documentclass[%
 aip,
 amsmath,amssymb,
 reprint,%
]{revtex4-1}

\usepackage{xcolor}
\usepackage{graphicx}
\usepackage{dcolumn}
\usepackage{bm}

\usepackage[utf8]{inputenc}
\usepackage[T1]{fontenc}
\usepackage{mathptmx}
\usepackage[textsize=footnotesize]{todonotes}
\usepackage{glossaries}
\usepackage{booktabs}
\usepackage{multirow}
\usepackage{float}

\usepackage[normalem]{ulem}
\usepackage{soul}

\newacronym{cmi}{CMI}{conditional mutual information}

\begin{document}

\preprint{AIP/123-QED}

\title{Detection of coupling in Duffing oscillator systems}

\author{Martin Brešar}%
\email{martin.bresar@ijs.si}
\affiliation{ 
Department of Systems and Control,
Jožef Stefan Institute,
Jamova cesta 39, SI-1000 Ljubljana, Slovenia
}%
\affiliation{ 
Jožef Stefan International Postgraduate School, Jamova cesta 39, SI-1000 Ljubljana, Slovenia
}
\author{Pavle Boškoski}%
\email{pavle.boskoski@ijs.si}
\affiliation{ 
Department of Systems and Control,
Jožef Stefan Institute,
Jamova cesta 39, SI-1000 Ljubljana, Slovenia
}%

\author{Martin Horvat}
\email{martin.horvat@fmf.uni-lj.si}
\affiliation{
Faculty of mathematics and physics, 
University of Ljubljana, 
Jadranska cesta 19, SI-1000 Ljubljana, Slovenia}


\begin{abstract}
In complex dynamical systems, the detection of coupling and its direction from observed time series is a challenging task. We study coupling in coupled Duffing oscillator systems in regular and chaotic dynamical regimes. By observing the conditional mutual information (CMI) based on the Shannon entropy, we successfully infer the direction of coupling for different system regimes. Moreover, we show that in the weak coupling limit the values of CMI can be used to infer the coupling parameters by computing the derivative of the conditional mutual information with respect to the coupling strength, called the {\it information susceptibility}. The complete numerical implementation is available at \url{https://repo.ijs.si/mbresar/duffing-cmi}.

\end{abstract}

\maketitle

\begin{quotation}
Complex systems encountered in nature can be divided into simpler subsystems that are coupled to each other, a prime example being a biological system. The problem of the detection of coupling from measured time series of subsystems has been of great interest in different branches of science. In this article, we study methods for the detection of coupling based on information transfer on a mathematical model. We show that information transfer estimation can be used to infer the exact value of the model coupling parameter.
\end{quotation}

\section{\label{sec:level1}Introduction}

In nature we encounter many complex systems that are divided into better known subsystems which are coupled to each other. The subsystems can be nonautonomous, i.e. time-dependent. In experimental practice, we typically only have measurements of observables of the identified individual subsystems in the form of signals. Studying these signals naturally raises a question of what we can infer from them about the system structure, i.e., subsystem properties and their coupling. In this paper, we focus predominantly on the last property.

More precisely, we study the effects of time-independent coupling between two nonautonomous systems through information transfer or flow between the trajectories of the systems determined by the conditional mutual information \cite{ThomasM.Cover2006}.  As an example of such systems, we use the well known Duffing oscillators \cite{Holmes1979}, i.e., periodically driven nonlinear harmonic oscillators. Although information flow is usually observed in the context of detecting the direction of coupling, the focus here is on the absolute values of information flow in a mathematical model. The goal is to solve the inverse problem of inferring the coupling parameter in different coupled nonautonomous oscillator systems from the calculated information flow.

Estimating the conditional mutual information has been successfully applied in numerous studies, including biological systems. For example, information flow has been used in the analysis of EEG recordings of epileptic patients~\cite{Palus2001}, in rats under the influence of different anaesthetics~\cite{Musizza2007}, and in the human cardiorespiratory system~\cite{Palus2007}. Studies have also been done on various theoretical systems, such as simple phase oscillators~\cite{Palus2003}, discrete Henon maps~\cite{Palus2001}, Rossler and Lorenz systems~\cite{Palus2007}, and different chaotic, stochastic, and nonautonomous systems~\cite{Clemson2014}. It has been shown that information flow can correctly detect both the existence and the direction of coupling in two coupled, but not yet synchronised systems. This means that the coupling must be sufficiently small. While it has been shown that the information flow between systems increases as the coupling increases, this relationship has not been thoroughly investigated.

Another measure for quantifying information transfer between systems is the transfer entropy proposed by \citet{Schreiber2000}. It has been shown that the transfer entropy in a specific (although generally useful) setup is mathematically equivalent to the conditional mutual information, see e.g. \cite[Appendix A]{Palus2007}. In a more general setup, the transfer entropy can suffer from a lack of statistical significance of distributions (due to high parameter dimensionality) used in calculations for short input signals.

First, we study the numerical properties of the information flow calculation, where the convergence was found to be faster than expected. Next, we consider the behavior of the information flow when different pairs of Duffing systems are coupled.
Addressing the issue of the dependence of the information flow on the coupling strength, we study the inverse problem, i.e., inferring the coupling parameter from the information flow. The result is an estimate of the information susceptibility as a derivative of the information flow. For details, see Section \ref{chapter infosusc}. It was found that the inverse problem is solvable in the limit of strong coupling and for most cases of driven chaotic systems in the limit of weak coupling.
Finally, the range of the coupling parameter where this is possible is given for individual pairs of systems.

\section{Inferring the direction of coupling}

Consider two nonautonomous dynamical systems with trajectories denoted by $\mathbf{x}(t)$ and $\mathbf{y}(t)$ and let the time evolution of such coupled systems be given by equations
\begin{align}
&\frac{d\mathbf{x}(t)}{dt} = \mathbf{f}_x(\mathbf{x},t) + \mathbf{g}_x(\mathbf{x},\mathbf{y},t) \\
&\frac{d\mathbf{y}(t)}{dt} = \mathbf{f}_y(\mathbf{y},t) + \mathbf{g}_y(\mathbf{y},\mathbf{x},t),
\end{align}
where $\mathbf{x}(t)$ and $\mathbf{y}(t)$ can be multidimensional and are therefore written in bold letters. In the case where both $\mathbf{g}_x$ and $\mathbf{g}_y$ are nonzero, the coupling is bidirectional. In our studies of the inverse problem the coupling functions ${\bf g}_x$ and ${\bf g}_y$ are time indepedent. If only one is nonzero, the coupling is unidirectional. In the subsequent analysis, only unidirectional coupling is analysed. When $\mathbf{g}_y=0$ and $\mathbf{g}_x \neq 0$, we call $\mathbf{x}(t)$ the driven system and $\mathbf{y}(t)$ the driving system.

The trajectories of both subsystems $\mathbf{x}(t)$ and $\mathbf{y}(t)$ are observed at equally spaced times $t_i = t_0 + i\Delta t$ for $i\in \mathbb{Z}$.  Thus, we obtain time series of these trajetories $(\mathbf{x}(t_i) : i \in \mathbb{Z})$ and  $(\mathbf{y}(t_i): i \in \mathbb{Z})$. The methods presented in the following are used to determine the direction and, in some cases, the magnitude of the coupling based on such measured time series.

\subsection{Conditional mutual information}

In the context of information theory, the amount of information being transferred from one system to another is quantified by the Shannon entropy \cite{ThomasM.Cover2006}. The Shannon entropy of a random variable $X$ with a probability distribution $p_X$ and its support $R_X$ (subset of the range of $X$) is defined as
\begin{equation}
\label{entropy1d}
H(X) = -\sum_{x \in R_X} p_X(x) \log p_X(x).
\end{equation}
The joint entropy of two random variables $X$ and $Y$ with a joint probability distribution $p_{X,Y}$ is defined as
\begin{equation}
H(X,Y) = -\sum_{x \in R_X, y \in R_Y} p_{X,Y}(x, y) \log p_{X,Y}(x, y).
\end{equation}
Their mutual information $I(X,Y)$ , which is a measure of the amount of common information contained in $X$ and $Y$, is defined as
\begin{equation}
I(X,Y) = H(X) + H(Y) - H(X, Y)
\end{equation}
and is maximal if $X = Y$ and equal to $0$ if $X$ and $Y$ are independent.
The conditional entropy of $X$ and $Y$ is
\begin{equation}
H(X|Y) = -\sum_{x \in R_X, y \in R_Y} p_{X,Y}(x, y) \log p_{X|Y}(x|y),
\end{equation}
where $p_{X|Y}$ is the conditional probability distribution.
Conditional entropy measures the uncertainty in $X$, provided that everything about $Y$ is known.
Conditional entropy $H(X, Y|Z)$ is defined similarly, which leads to \gls{cmi}
\begin{equation}
\label{CMI1}
I(X,Y|Z) = H(X|Z) + H(Y|Z) - H(X,Y|Z).
\end{equation}
It quantifies the average amount of common information contained in $X$ and $Y$ given the value of $Z$.

While a measurement almost certainly contains uncorrelated noise that is stochastic in nature, the underlying dynamics of the measured system can still be deterministic.

Let us consider two possibly coupled systems with trajectories $\mathbf{x}(t)$ and $\mathbf{y}(t)$. Their corresponding time series are $(\mathbf{x}(t_i) : i \in \mathbb{Z})$ and $(\mathbf{y}(t_i) : i \in \mathbb{Z})$. Their values are distributed according to the probability distributions $p_\mathbf{X}$ and $p_\mathbf{Y}$. We think of the values in the series $\mathbf{x}(t_i)$ and $\mathbf{y}(t_i)$ as a realization of the random variables denoted by $\mathbf{X}$ and $\mathbf{Y}$ at time $t_i$. By just discussing their distributions, we lose the information about time. To restore the notion of time, we need to introduce the time-lagged random variables $\mathbf{X}_\tau$ and $\mathbf{Y}_\tau$ with realizations ${\bf x}(t_{i + \tau})$ and ${\bf y}(t_{i + \tau})$ at times $t_i$ and discuss them in parallel with their non time-lagged versions.

In order to infer a possible influence of the trajectory $\mathbf{x}(t)$ on $\mathbf{y}(t)$ from their time series, we need to compute the mutual information between the random variable of the first system $\mathbf{X}$ and the time-lagged random variable of the second system $\mathbf{Y}_\tau$ given by $I(\mathbf{X}, \mathbf{Y}_{\tau})$.  However, the mutual information captures the connection between $\mathbf{X}$ and $\mathbf{Y}$ as well as between $\mathbf{Y}$ and $\mathbf{Y}_{\tau}$ when the two systems are not independent. Therefore, \citet{Palus2001} proposed to observe the conditional mutual information $I(\mathbf{X}, \mathbf{Y}_{\tau}|\mathbf{Y})$ to infer causal relations between trajectories $\mathbf{x}(t)$ and $\mathbf{y}(t)$. This quantity only captures the net information flow from $\mathbf{x}(t)$ to $\mathbf{y}(t)$. A tweak to increase precision is to instead calculate
\begin{equation}
\label{CMI2}
I(\mathbf{X}, \Delta_{\tau}\mathbf{Y}|\mathbf{Y}) = H(\mathbf{X}|\mathbf{Y})+H(\Delta_{\tau}\mathbf{Y}|\mathbf{Y}) - H(\mathbf{X},\Delta_{\tau}\mathbf{Y}|\mathbf{Y}),
\end{equation}
where $\Delta_{\tau}\mathbf{Y} = \mathbf{Y}_\tau  - \mathbf{Y}$. This allows us to look at small local changes in $\mathbf{y}(t)$ rather than the entire range of data.
Information flow from $\mathbf{y}(t)$ to $\mathbf{x}(t)$ is calculated as $I(\mathbf{Y}, \Delta_{\tau}\mathbf{X}|\mathbf{X})$ by swapping $\mathbf{X}$ and $\mathbf{Y}$.

\subsection{Information flow}

Both $I(\mathbf{X},\mathbf{Y}_{\tau}|\mathbf{Y})$ and $I(\mathbf{X},\Delta_{\tau}\mathbf{Y}|\mathbf{Y})$ are zero for $\tau=0$.
However, $\lim_{\tau \rightarrow 0^+}I(\mathbf{X},\mathbf{Y}_{\tau}|\mathbf{Y}) = 0$ and $\lim_{\tau \rightarrow 0^+}I(\mathbf{X},\Delta_{\tau}\mathbf{Y}|\mathbf{Y}) \neq 0$.
Since we are interested in small time lags in Section \ref{cmidependences} and  \ref{section inverse}, we use the second definition.

To obtain a numerically more robust measure, we can average the \glspl{cmi} $I(\mathbf{X}, \Delta_{\tau}\mathbf{Y}|\mathbf{Y})$ over some time lags $\tau$ in order to better approximate the information flow between two systems \cite{Palus2001}.
\begin{equation}
\label{inflow}
I(\mathbf{x} \rightarrow \mathbf{y}) = \frac{1}{l}\sum_{i = 1}^{l} I(\mathbf{X}, \Delta_{i}\mathbf{Y}|\mathbf{Y})
\end{equation}
The appropriate values of $l$ are discussed in Section \ref{section lag dependence}.

When calculating information flow from one subsystem to another, we use the values of trajectories. The trajectories capture the full extend of information about the dynamics of a particular system and can be multidimensional. In realistic situations, however, the measurement of a system or a subsystem is usually one-dimensional. A possible solution to account for a multidimensional phase space is the use of time-delay embedding vectors \cite{Rand1981} or instantaneous phase representation in the case of oscillatory systems \cite{Rosenblum1996, Palus2003}.

\section{Numerical analysis}

Information theory is thoroughly researched and well understood. However, numerical computation of entropies is challenging due to limitations and necessary approximations. It is therefore important to be aware of them and study their effect on known systems in order to be able to analyze unknown ones.

\subsection{Numerical implementation}
\label{Numerical implementation}

A convenient way to numerically implement the calculation of \gls{cmi} is to use the chain rule for entropies $H(X|Y) = H(X,Y) - H(Y)$ and rewrite \eqref{CMI1} as
\begin{equation}
\label{cminum}
I(\mathbf{X}, \Delta_{\tau}\mathbf{Y}|\mathbf{Y}) = H(\mathbf{X}, \mathbf{Y}) + H(\Delta_{\tau}\mathbf{Y}, \mathbf{Y}) - H(\mathbf{Y}) - H(\mathbf{X}, \Delta_{\tau}\mathbf{Y},\mathbf{Y}) \>.
\end{equation}
The biggest issue in calculating \gls{cmi} \eqref{cminum} from the involved entropies is the estimation of the underlying probability distributions by the corresponding histograms. The latter are computed from long enough series of trajectories: having a time series of length $n$, we assign its values in each dimension into $N$ bins, where $N\ll n$, and construct a histogram. Multidimensional histograms are used to estimate multidimensional probability distributions. Consider a system with two $k$-dimensional subsystems. When computing the joint entropy in \gls{cmi} \eqref{cminum}, the histogram used in its calculation has $N^{3k}$ bins. Individual bins must contain a significant number of data points to sufficiently approximate the distribution. Therefore, the total number of bins cannot be too large.

The simplest way to construct a histogram is to use equidistant binning, which means that the widths of all histogram bins are equal. We found this binning procedure to be sufficient for our problem. Other binning procedures can also be used, such as marginal equiquantization, where the marginal bins are chosen so that they each contain the same number of data points \cite{Palus2007}.

\subsection{Duffing oscillator systems}

The model we have chosen for numerical analysis is the Duffing oscillator system, which is studied in detail in \cite{Kovacic2011}.
Its properties make it very convenient, since it is a nonautonomous oscillator that can be in a regular or in a chaotic dynamical regime.
It is defined by
\begin{equation}
\label{duffing}
\frac{d^2x}{dt^2} + \delta \frac{dx}{dt} + \alpha x + \beta x^3 = \gamma \cos(\omega t).
\end{equation}
It is a nonlinear periodically forced oscillator with damping. Its trajectory is denoted by  $(x(t), \frac{dx}{dt}(t))$ and the associated phase space is two-dimensional. The choice of the model parameters $(\delta, \alpha, \beta, \gamma, \omega)$ yields different systems which can be in a regular or in a chaotic dynamical regime \cite{Ueda1979}. This allows us to test the behavior of \gls{cmi} on systems in different regimes.

\begin{figure}
\includegraphics[width = \linewidth]{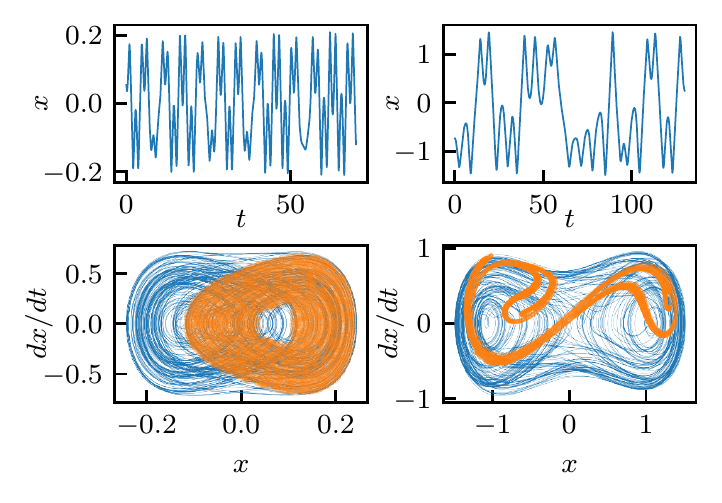}
\caption{Chaotic Duffing oscillator systems $C_1$ (left) and $C_4$ (right). The orange dots represent a stroboscopic map at $\cos(\omega t) = 1$ obtained for long times $t_{max} = 10^5$.}
\label{exampleduffing}
\end{figure}

For regular behavior, a high damping $\delta = 0.2$ and a linear potential $\beta = 0$ were chosen. The orbits of the chosen regular systems are circles, same as in normal driven harmonic oscillators, while chaotic systems exhibit richer dynamics. For chaotic behavior, it is beneficial to have low damping. In fact, increasing the damping parameter $\delta$ of a chaotic system makes it regular. Four different chaotic systems denoted $C_1$ to $C_4$ and two regular ones denoted $R_1$ and $R_2$ were chosen. The parameters and the maximal Lyapunov exponents of the chosen systems are presented in \tablename~\ref{tableduffing}. The maximal Lyapunov exponents were calculated from the stroboscopic map at $\cos{\omega t} = 1$ using the Benettin algorithm \cite{Benettin1980}. The negative maximal Lyapunov exponents of regular systems were obtained due to orbits converging towards a limit cycle. Two examples of chaotic systems are shown in \figurename~\ref{exampleduffing}.

\begin{table}
\caption{\label{tableduffing} Parameters of the six Duffing systems under consideration and their maximal Lyapunov exponents $\lambda_{\rm max}$. The systems $R_1, R_2$ are in a regular regime. The chaotic systems $C_1,C_2$ have a high nonlinear term and $C_3, C_4$ have a low nonlinear term.}
\begin{ruledtabular}
\begin{tabular}{lcccccr}
System & $\delta$ & $\alpha$ & $\beta$ & $\gamma$ & $\omega$ & $\lambda_{\rm max}$\\
\hline
$R_1$ & 0.2 & 1 & 0 & 1 & 0.51 & -1.23(2)\\
$R_2$ & 0.2 & 1 & 0 &1 & 1.83 & -0.34(3)\\
$C_1$ & 0 & 1 & 400 & 1 &1.1 & 0.6(2)\\
$C_2$ & 0 & 1 & 80 & 1 & 3.1 & 0.2(3)\\
$C_3$ & 0.3 & -1 & 1 & 0.5 & 1.2 & 0.5(7)\\
$C_4$ & 0.2 & -1 & 1 & 0.3 & 1 & 0.9(9)\\
\end{tabular}
\end{ruledtabular}
\end{table}

Systems $C_1$ and $C_2$ are similar to each other, and so are $C_3$ and $C_4$, as can be seen from the table of parameters. Systems $C_1$ and $C_2$ have a highly nonlinear potential (high $\beta$) and no damping ($\delta=0$), while $C_3$ and $C_4$ have a small $\beta$ and a high $\delta$, making their dynamics less "wild" but still chaotic. In fact, the system $C_4$ has a fractal attractor with dimension of around $1.4$ \cite{Tarnopolski2013}. The driving frequencies $\omega$ of the different systems were chosen so that one is not a multiple of another.

\subsection{Duffing system's CMIs}
Two Duffing systems can be coupled in different ways. We have chosen a unidirectional linear coupling of the form $+\epsilon_1 (x_2 - x_1)$ due to its intuitive interpretation.
Such a coupling forces the driven system towards the driving system and can cause synchronization.
If the coupling is very strong, the systems become identical.
Coupled Duffing systems are thus defined as:
\begin{align}
\label{eqcoupduff1}
&\frac{d^2x_1}{dt^2} + \delta_1 \frac{dx_1}{dt} + \alpha_1 x_1 + \beta_1 x_1^3 = \gamma_1 \cos(\omega_1 t) + \epsilon_1(x_2 - x_1) \\
&\frac{d^2x_2}{dt^2} + \delta_2 \frac{dx_2}{dt} + \alpha_2 x_2 + \beta_2 x_2^3 = \gamma_2 \cos(\omega_2 t).
\label{eqcoupduff2}
\end{align}
While the term driven can also be used to describe an individual Duffing oscillator, i.e., it is driven by an external periodic force, we use the terms driving and driven to distinguish between $x_1$ and $x_2$. We say that the driven system $x_1$ depends on the driving system $x_2$.

We write a trajectory of the coupled Duffing system as $(x_1(t), \frac{dx_1}{dt}(t), x_2(t), \frac{dx_2}{dt}(t))$ and trajectories of its subsystems as ${\bf x}_i(t) = (x_i(t),  \frac{dx_i}{dt}(t))$ for $i=1,2$.  The phase space of the whole system is thus four-dimensional. In our calculations we ensured that our system is on the attractor, so that the results do not depend on the initial conditions. This was accomplished by taking the initial conditions $(x_1(0), \frac{dx_1}{dt}(0), x_2(0), \frac{dx_2}{dt}(0))= (0, 0, 0, 0)$ and only considering the data from the trajectories from time $t = 100$ onward. This way, the energy of the system is gained entirely from the nonautonomous parts of subsystems \eqref{eqcoupduff1} and \eqref{eqcoupduff2}.

The trajectory of the whole system, or more precisely, the corresponding time series, was obtained by first transforming the equations of motion \eqref{eqcoupduff1} and \eqref{eqcoupduff2} to a system of four first order differential equations and then integrating them using the Runge-Kutta 4 integrator with a fixed integration step equal to $\Delta t = 0.002$ and the time series length of $n = 10^7$ throughout our study. The use of a constant integration step is convenient for calculating probability distributions of points on the trajectory. In the calculation of the histogram of points pertaining to the evaluation of \gls{cmi}, we use equidistant binning for each two-dimensional subsystem, as explained in Section \ref{Numerical implementation}. In this study we use $4$ bins for $x$ direction and $4$ for $dx/dt$ direction unless otherwise specified. This effectively gives $16$ bins per subsystem.

\begin{figure}
\includegraphics[width = \linewidth]{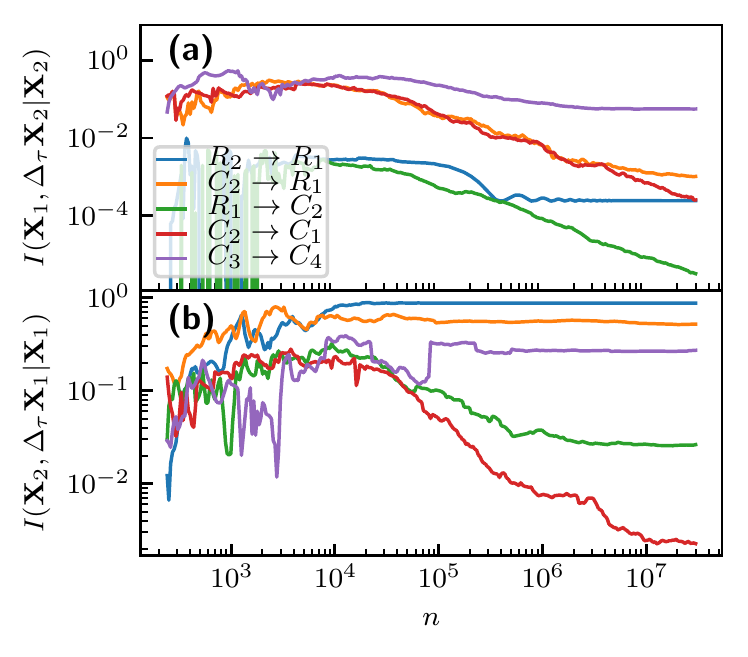}
\caption{The convergence of \glspl{cmi} \eqref{CMI2} for time lag $\tau= 1$ between different systems in the direction without coupling (a) and in the direction of coupling (b). The coupling parameter is $\epsilon_1 = 0.5$, the integration step is $\Delta t = 0.002$ and $4$ bins per scalar variable were used in the histograms.}
\label{convergence}
\end{figure}

The convergence of \gls{cmi} with the length of the trajectory $t$, more precisely the number of steps in the sample $n (\approx t/\Delta t)$, is shown in \figurename~\ref{convergence}. In the direction without coupling (a), the \gls{cmi} values decrease with the length of the time series $n$ and stabilise at a nonzero value. We attribute this to approximation errors of the discrete probability distributions. In the direction of coupling (b), the values converge to a constant. The convergence is slower in the presence of chaotic systems, especially for $C_1$ and $C_2$. In a certain range of $\epsilon_1$ we see that the dependence is linear in the log-log scale. Fitting a function of the form $f(t) = A t^{-C}$ in this range, we obtain the coefficient $C = 1$ for convergence, even for chaotic systems.

This is a bit surprising, since CMI is essentially a difference of entropies determined by the empirical distributions which converge for periodic and stochastic (chaotic) signals as $O(t^{-1})$ and $O(t^{-1/2})$, respectively. The latter was checked in our study. The $O(t^{-1})$ convergence of CMI is therefore faster than of its ingredients. This seems to be a consequence of the fact that the stochastic corrections to entropies of order of magnitude $O(t^ {-1/2})$ are correlated and cancel each other out in the calculation of CMI, and only the next possible order $O (1/ t)$ remains. This is good news for the usability of CMI, as even shorter samples can deliver relatively precise results. Further discussion of the convergence is given in Appendix \ref{appendix convergence}.

\bigskip
 
\section{Analysis of CMI and information flow}\label{cmidependences}

\begin{figure*}
\includegraphics[width = \linewidth]{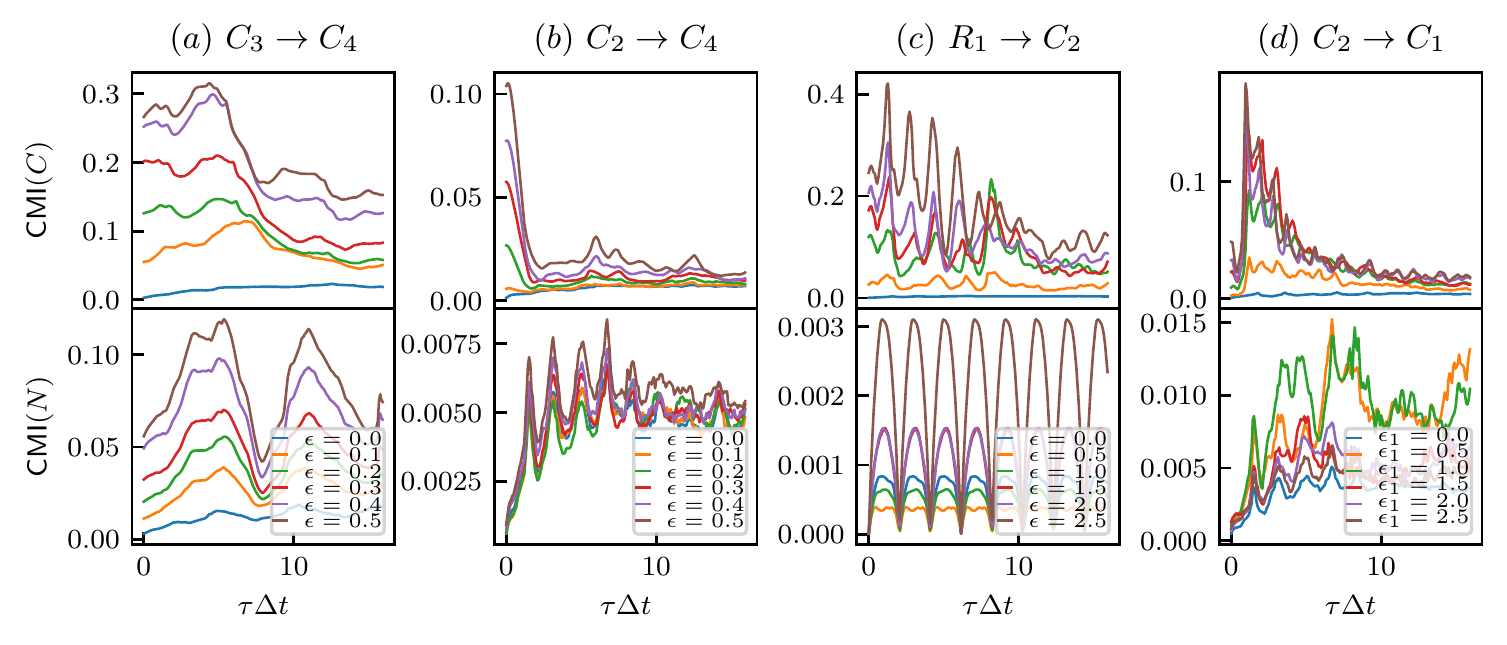}
\caption{The dependence of \gls{cmi} in both directions on the time lag $\tau$. Four different systems are examined from (a) to (d).}
\label{lagdependence}
\end{figure*}

We have defined \gls{cmi} between two subsystems and information flow as the average of \glspl{cmi} over time lags. In this section, we study the behavior of these quantities for various combinations of coupled Duffing systems. Mainly, we examine the dependence of \gls{cmi} on the time lag and the dependence of information flow on the coupling parameter. We denote the direction of coupling by $(C)$ and the direction without coupling by $(N)$, e.g., information flow in the direction of coupling is denoted by $I(C)$. Here, $C$ stands for coupled and $N$ stands for not coupled. 

\subsection{Dependence of \gls{cmi} on the time lag}
\label{section lag dependence}

First, it is important to observe the dependence of \glspl{cmi} \eqref{CMI2} on the time lag $\tau$ in order to determine the averaging in the definition of information flow \eqref{inflow}. It is shown for four examples of coupled Duffing systems in \figurename~\ref{lagdependence}.

Three main conclusions can be drawn from these dependences. First, we note that the values in the direction of coupling \gls{cmi}$(C)$ (upper row) are much larger than \gls{cmi}$(N)$ (lower row), as can be seen in all examples from (a) to (d). Therefore, the \gls{cmi} values can be used to determine the direction of coupling. Second, we see that \gls{cmi}$(C)$ increases with increased coupling $\epsilon_1$. This means that our measure of coupling between systems increases with coupling. Finally,  \gls{cmi}$(C)$ decreases with increased time lag.
It falls with the correlation time of the time-lagged system. On the contrary, \gls{cmi}$(N)$ increases with increased time lag. An exception is the system $C_2\rightarrow C_1$, where \gls{cmi}$(C)$ increases again to an even higher value at $\tau \Delta t \approx 1$ and then proceeds to decrease.

This shows that the appropriate averaging in the definition \eqref{inflow} is over small time lags $\tau$, where the values of \gls{cmi} are large and information about the driving system is still present in the driven one. The averaging decreases errors due to dynamic fluctuations and possible noise.
The simulation of the system is noiseless and the errors of dynamical fluctuations are decreased with the length of the time series. Therefore, taking only one value of \gls{cmi} is shown to be sufficient. Since \gls{cmi}$(C)$ generally falls and \gls{cmi}$(N)$ generally increases for small time lags, the values at the smallest possible time lag $\tau = 1$ are taken into account. Our measure of information flow in either direction is therefore the very first point in \figurename~\ref{lagdependence}.

Since oscillator systems are similar to themselves when they are time-lagged, we also detect oscillations at the frequency of the time-lagged system. They are most expressed for the system $R_1 \rightarrow C_2$. It is important to mention that while \gls{cmi} generally decreases with $\tau$ for small coupling, the dependence can be different for bigger coupling.

\subsection{Dependence of information flow on the coupling parameter} \label{section epsdependence}

\begin{figure*}
\includegraphics[width = \linewidth]{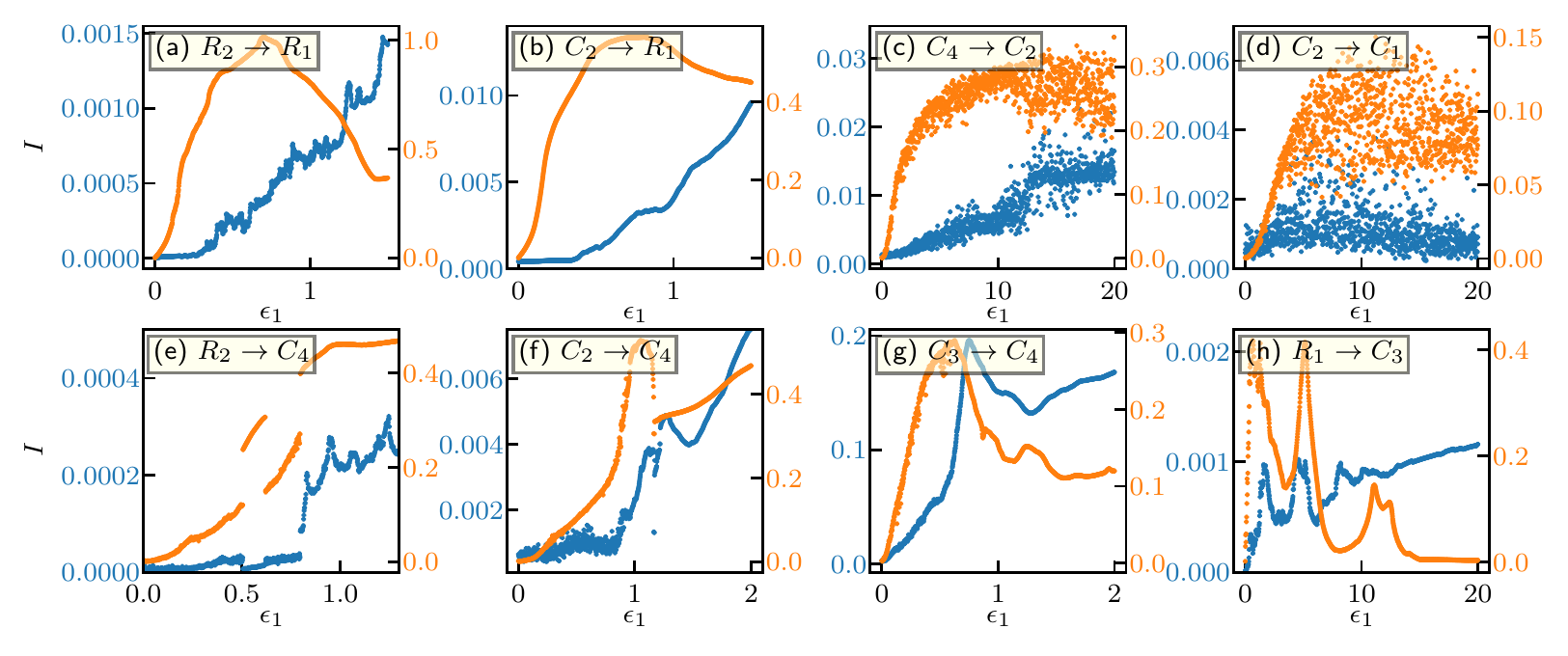}
\caption{The dependence of the information flow \eqref{inflow} on the coupling parameter $\epsilon_1$. The orange lines represent the flow in the direction of coupling $I(C)$ and the blue lines in the direction without coupling $I(N)$. The dependences for weak coupling are shown for various systems in (a) to (g). For the system $R_1 \rightarrow C_3$, strong coupling is also shown in (h). The calculations were done for $1000$ evenly spaced values of $\epsilon_1$ for each case.}
\label{epsdependence}
\end{figure*}

We can now study the information flow between different systems to see how it characterises them.
The focus here is on observing the dependence of the information flow on the coupling parameter.
This dependence is shown in \figurename~\ref{epsdependence} for various coupled Duffing systems.

The general behavior is similar for any two coupled systems.
The information flow $I(C)$ is very close to $0$ in the absence of coupling, since the systems are independent.
It then generally increases as the coupling parameter increases.
This can be seen in the examples from (a) to (f) in \figurename~\ref{epsdependence}.

For intermediate values, this dependence can have many local extremes until it begins to decrease and reaches zero in the limit $\epsilon_1 \rightarrow \infty$, as indicated by \figurename~\ref{epsdependence} (h).
In this limit, the systems are the same because the orbit of one system is pushed towards that of the other, and there is no information transfer between identical systems.

It can also be seen that $I(C)$ is much greater than $I(N)$.
This means that the direction of coupling can be easily inferred from the information flow.
There are some exceptions when it comes to bigger coupling, one of which can be seen in \figurename~\ref{epsdependence} (g).
We have not found a general rule for which systems exhibit this property. However, it seems to occur when similar pairs of systems are coupled, for example for all four cases $C_i \rightarrow R_j$, for  $i, j \in \{1, 2\}$.

In the limits of weak and strong coupling, the dependence can be monotonic, which will be discussed in detail in the next section.
A difference is observed between chaotic systems with high and with low nonlinear terms $\beta$. If the driven system is one with a high nonlinear term $C_1$ or $C_2$, the dependence appears much more noisy, especially when two of these systems are coupled. Monotonicity is harder to show or find in these cases, which makes solving the inverse problem harder.

\section{The inverse problem}
\label{section inverse}

In this section, we are interested in whether the coupling parameter is uniquely determined by the information flow $I(C)$ in some range of values, i.e., whether this dependence is monotonic. If this is the case, the coupling parameter can be inferred from the time series of two known systems by calculating the information flow.

For most pairs of systems, there are two cases that exhibit monotonic dependence of the information flow on the coupling parameter, as indicated in Section \ref{section epsdependence}. The first is the limit of strong coupling, i.e., for large values of the coupling parameter. The second is the limit of weak coupling, which is most often observed in nature. These limiting cases will be studied in detail. For intermediate values, the inverse problem is generally not solvable.

\subsection{Limit of strong coupling}

\begin{figure}
\includegraphics[width = \linewidth]{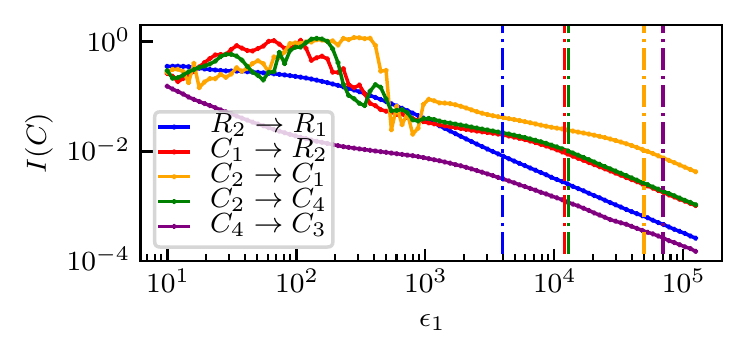}
\caption{Information flow in the direction of coupling for large $\epsilon_1$ for different coupled systems. Vertical lines mark the lower boundaries of areas used for regression.}
\label{inverselimit2}
\end{figure}

First, we consider the limiting case of large coupling parameters $\epsilon_1$.
As explained in Section \ref{section epsdependence}, the information flow $I(C)$ decreases towards zero in this limit.
The dependence for large $\epsilon_1$ is shown in \figurename~\ref{inverselimit2} in double logarithmic scale.
Since the dependence in this scale is linear for sufficiently large $\epsilon_1$, we know its form is
\begin{equation}
f(\epsilon_1) = A\epsilon_1^{-B}.
\end{equation}
Using least squares regression on points in the marked areas, the value $B=1$ was obtained for all cases.
The obtained formulas hold for a wide range of values of $\epsilon_1$.
The value $B=1$ holds regardless of the number of bins chosen, as does the range of $\epsilon_1$ with this dependence (checked up to $24$ bins per variable).
Thus, the inverse problem is solvable in the limit of large coupling.

\subsection{Limit of weak coupling}

The case of weak coupling is much more interesting, since it is usually encountered in physical systems. 
All numerical tests suggest that in most cases the information flow grows monotonically with increasing $\epsilon_1$, up to a value $\epsilon_{max}$ where the monotonicity ends.
It may end due to a distinct peak in the dependence, as in \figurename~\ref{epsdependence} (f),(g), due to a discrete jump as in \figurename~\ref{epsdependence} (e), or due to the appearence of noise as in \figurename~\ref{epsdependence} (c).
The analysis of different coupled Duffing systems gives different answers to the solvability of the inverse problem.

\begin{figure}
\includegraphics[width = \linewidth]{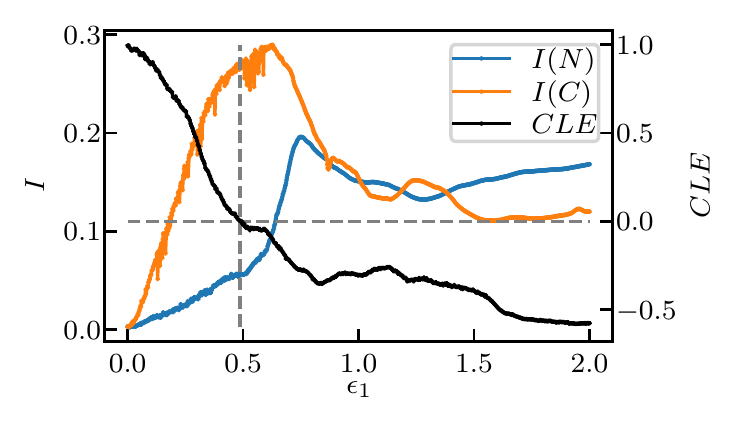}
\caption{Information flow in both directions and the largest Lyapunov exponent of the driven subsystem (conditional Lyapunov exponent) for system $C_3 \rightarrow C_4$. Gray lines represent the point where the exponent reaches zero, i.e., the point of synchronization.}
\label{figlyapunov}
\end{figure}

In most cases where the monotonicity ends with a distinct peak in the dependence on the coupling parameter, it is due to the synchronization of subsystems. In a mathematical model, the appearance of synchronization can be observed with conditional Lyapunov exponents (CLE), i.e., the Lyapunov exponents of the driven subsystem \cite{Pecora2015}. The dependence of the information flow on $\epsilon_1$ and the dependence of CLE on $\epsilon_1$ is shown in \figurename~\ref{figlyapunov} for one of these cases. Synchronization, which is marked by the largest exponent becoming negative, occurs close to the point where the dependence of the information flow suddenly changes, which is similar to observations made in \cite{Palus2007, Palus2018}. A thorough survey on synchronization is given in \cite{synchronization2001}.

In cases of the discrete jumps in the information flow dependence, the observation of phase state portraits and conditional Lyapunov exponents reveal bifurcations of the driven subsystem.
These cases are examined in detail in Appendix \ref{appendix systems}.

Let us now consider two coupled regular systems $R_2 \rightarrow R_1$.
The dependence in \figurename~\ref{epsdependence} (a) indicates a monotonic trend up to $\epsilon_{max} \approx 0.7$.
This value should not strongly depend on the details of numerical calculation in order to determine the value of $\epsilon_{max}$ as a property of a system.

\begin{figure}
\includegraphics[width = \linewidth]{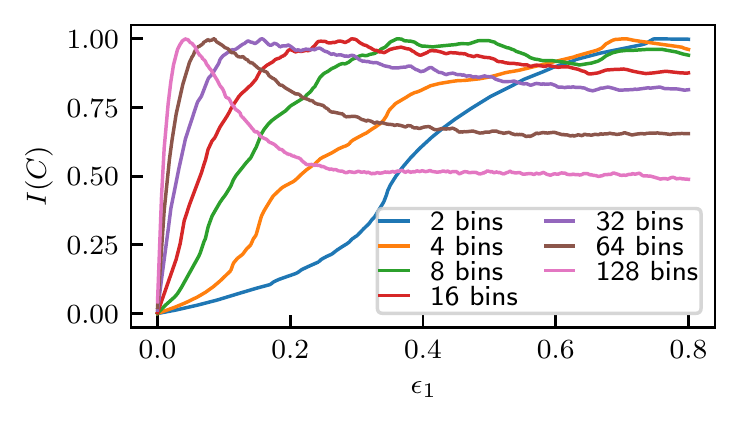}
\caption{Information flow in the direction of coupling for unidirectionally coupled systems $R_2 \rightarrow R_1$, calculated with different numbers of bins per variable $N$. The dependences are normalized to be easily comparable.}
\label{bindependenceRR}
\end{figure}

By increasing the number of marginal bins $N$, the results approach the exact analytic dependence for smooth variables \cite{Hardle2004}.
However, $\epsilon_{max}$ moves towards $0$ as the number of marginal bins of the histograms increases, as can be seen in \figurename~\ref{bindependenceRR}.
This is observed in all cases when the driven system is regular.
We have done the same test with two coupled harmonic oscillators and obtained very similar results.
Therefore, we suspect that this is due to the singularity of the probability distribution $p(x,dx/dt)$ of regular systems, since their orbits are circles in 2D space and even a very weak coupling changes the probability distributions significantly so that it cannot be explained by linear perturbation theory.

\begin{figure}
\includegraphics[width = \linewidth]{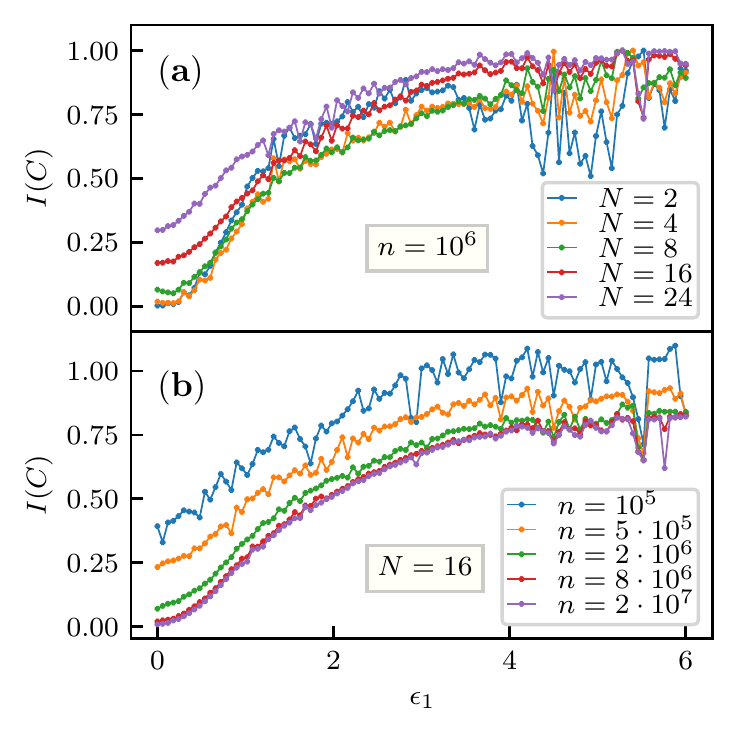}
\caption{Information flow in the direction of coupling for unidirectionally coupled systems $R_1 \rightarrow C_2$, calculated with data length $n = 10^6$ at different number of bins $N$ (a) and with $N=16$ at different $n$ (b). In (a), the dependences are normalized to be easily comparable.}
\label{bindependenceRC}
\end{figure}

For driven chaotic systems, the value of $\epsilon_{max}$ did not change for different numbers of histogram bins.
The numerical analysis of $R_1 \rightarrow C_2$ is done in \figurename~\ref{bindependenceRC}.
Increasing the number of marginal bins $N$ while keeping the length of the time series $n$ constant does not affect the general shape, as seen in \figurename~\ref{bindependenceRC} (a).
However, the values increase at $\epsilon_1 = 0$ because the bins become poorly populated. With increasing length $n$ at a high $N=16$, the value decreases again towards $0$, as seen in \figurename~\ref{bindependenceRC} (b). This shows that while a finer partition is generally beneficial, taking too many bins will cause them to be poorly populated and distorts the results.

Moreover, the dependence is found to converge towards a monotonic dependence as $n$ is further increased. Due to numerical imprecisions, the dependence will never be exactly monotonic when considering values of $\epsilon_1$ that are very close to each other. Instead, we show monotonicity empirically by doing ever more precise calculations. These can be done by increasing $n$, and perhaps $N$. Decreasing the time step of the integration $\Delta t$ changes the results only minimally.

For most cases of driven chaotic systems, monotonicity has been shown in the regime of weak coupling strength $\epsilon_1$.
For the coupling of two systems with high nonlinear term $C_i \rightarrow C_j$, for $i,j \in\{1,2\}$, seen in \figurename~\ref{epsdependence} (d), the dependence always contains a lot of noise that does not decrease with longer time series.
This is due to bifurcations already at very small values of the coupling parameter. Bifurcations of the driven system destroy the monotonicity, as explained in the next section.

\subsection{Information susceptibility}
\label{chapter infosusc}

\begin{table*}
\caption{Values of $\epsilon_{max}$ for various driven and driving systems. Cases where the monotonicity of the information flow in the low coupling regime could not be shown are marked with a question mark. Systems that exhibit non-continuous dependencies due to bifurcations are marked from \ref{appendixR2vC1} to \ref{appendixC1vC3} as a reference to the appendix where they are examined.}
\label{tableinverse}

\centering
\begin{tabular}{cc|c|c|c|c|c|c|l}
\cline{3-8}
& & \multicolumn{6}{ c| }{Driving system} \\ \cline{3-8}
& & $R_1$ & $R_2$ & $C_1$ & $C_2$ & $C_3$ & $C_4$\\ \cline{1-8}
\multicolumn{1}{ |c  }{\multirow{4}{*}{Driven system} } &
\multicolumn{1}{ |c| }{$C_1$} & $\epsilon_{max} = 1.(0)$ & $\epsilon_{max} = 0.3(8)^{\ref{appendixR2vC1}}$ & \ & $?$ & $\epsilon_{max} = 1.(9)$ & $\epsilon_{max} = 1.(8)$ & \\ \cline{2-8}
\multicolumn{1}{ |c  }{}                        &
\multicolumn{1}{ |c| }{$C_2$} & $\epsilon_{max} = 3.(9)$ & $\epsilon_{max} = 5.(4)$ & $?$ & \ & $\epsilon_{max} = 0.(9)$ & $\epsilon_{max} = 0.(7)$ &  \\ \cline{2-8}
\multicolumn{1}{ |c  }{} &
\multicolumn{1}{ |c| }{$C_3$} & $\epsilon_{max} = 0.0(4)$ & $\epsilon_{max} = 0.02(1)^{\ref{appendixR2vC3}}$ & $\epsilon_{max} = 0.02(3)^{\ref{appendixC1vC3}}$ & $\epsilon_{max} = 0.02(1)^{\ref{appendixC2vC3}}$ & \ & $\epsilon_{max} = 0.7(4)$ &  \\ \cline{2-8}
\multicolumn{1}{ |c  }{}                        &
\multicolumn{1}{ |c| }{$C_4$} & $\epsilon_{max} = 0.3(2)^{\ref{appendixR1vC4}}$ & $\epsilon_{max} = 0.5(1)^{\ref{appendixR2vC4}}$ & $\epsilon_{max} = 0.4(7)$ & $\epsilon_{max} = 1.0(4)$ & $\epsilon_{max} = 0.5(8)$ & \ &  \\ \cline{1-8}
\end{tabular}
\end{table*}

To determine the values of $\epsilon_{max}$, we introduce a smooth function $I_s$ obtained by fitting splines to the dependence of the information flow $I(C)$ on the coupling constant.
This allows us to calculate the information susceptibility, which we define as
\begin{equation}
\label{infosusc}
\chi_I(\epsilon_1) = \frac{dI_s}{d\epsilon_1}(\epsilon_1).
\end{equation}
The smooth dependence allows a simple calculation of the derivative, which is not possible for discrete points of the information flow.
The information susceptibility $\chi_I(\epsilon_1)$ is positive for small $\epsilon_1$. While it is positive, the dependence of \gls{cmi} on $\epsilon_1$ is monotonic and the inverse problem is solvable.
The value of $\epsilon_{max}$ is then marked by the change of sign of $\chi_I(\epsilon_1)$, i.e., the first point at which it is equal to $0$.

This was done for systems where a monotonic dependence on the coupling parameter was found. The function \texttt{UnivariateSpline} from the package \texttt{scipy.interpolate} in \texttt{Python 3} was used to obtain $I_s$. Parameters chosen were $k=3$ for cubic splines and a smoothing factor between $0.005 \leq s \leq 0.1$.

\begin{figure}
\includegraphics[width = \linewidth]{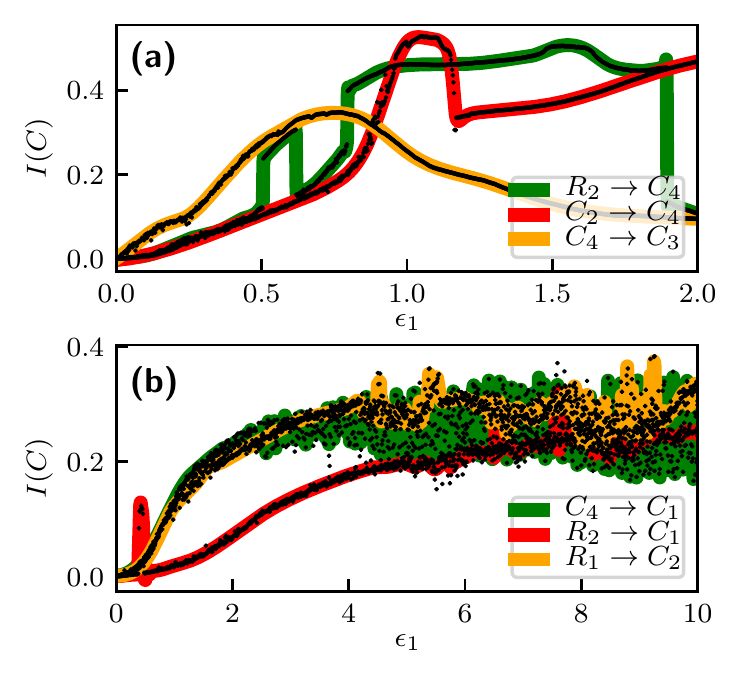}
\caption{Information flow data and their smoothed fit as a function of the coupling parameter $\epsilon_1$ for different pairs of systems, driven system having a small (a) and a large (b) nonlinear term.}
\label{fittingsplines}
\end{figure}

For the cases where the driven system is either $C_3$ or $C_4$, this is straightforward most of the times, as can be seen in \figurename~\ref{fittingsplines} (a).
An odd example $R_2 \rightarrow C_4$ can also be seen. It contains a nonsmooth dependence at $\epsilon_1 = 0.51$. Observing the phase space portraits of the driven system has shown that this is due to bifurcations. For $0.51 <\epsilon_1 <0.61$ and for $\epsilon_1 >0.8$ the driven subsystem $C_4$ is in a regular periodic regime. If the driven system's dynamics change significantly at some value of $\epsilon_1$ before a distinct peak in the information flow is observed (in this case at $\epsilon_1 = 0.51$), this value marks $\epsilon_{max}$. Such cases are shown in Appendix \ref{appendix systems}. A peak in the information flow can still be seen in such cases, but does not mark $\epsilon_{max}$.

For the cases seen in \figurename~\ref{fittingsplines} (b) where either $C_1$ or $C_2$ is the driven system, the exact value of $\epsilon_{max}$ could not be determined due to a noisy dependence already seen in \figurename~\ref{epsdependence} (c). When these two systems are driven, their orbits can get stuck in a region of phase space for some time. This can have a seemingly random effect on the information flow, which results in the noisiness. The monotonicity could only be reliably shown up to a certain point where this noise started. In these cases, this point was taken as $\epsilon_{max}$.

\begin{figure}
\includegraphics[width = \linewidth]{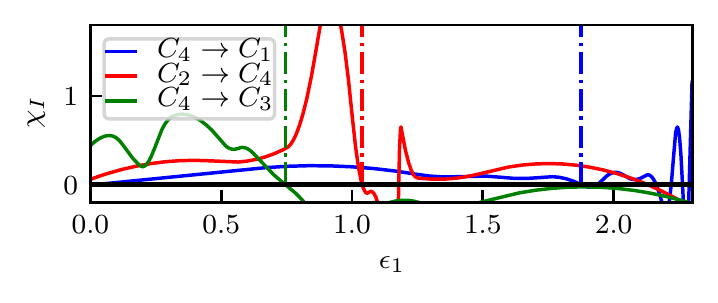}
\caption{Information susceptibility \eqref{infosusc} for different systems. Dashed vertical lines mark the obtained values of $\epsilon_{max}$.}
\label{figinfosusc}
\end{figure}

Once a smooth function is obtained, the derivative in \eqref{infosusc} is simple to calculate. Some examples of the information susceptibility and the obtained values of $\epsilon_{max}$ are shown in \figurename~\ref{figinfosusc}. The values of $\chi_I$ are always smaller in the cases of driven $C_1, C_2$ than for driven $C_3,C_4$, implying that these systems are less adaptable to the driving system.

Finally, we show the results for the solvability of the inverse problem in the limit of weak coupling in \tablename~\ref{tableinverse}. Only driven chaotic systems are considered, since the results for driven regular systems depend strongly on the partition of the phase space. Some properties of the individual systems and the information flow can be derived from the table.

Most notably, a difference between chaotic systems with a large and with a small nonlinear term is seen. For driven chaotic systems with a large nonlinear term, $\epsilon_{max}$ was mostly marked by the appearence of noise. Nevertheless, the values of $\epsilon_{max}$ are generally larger than for driven $C_3$, $C_4$. This means that $C_1$ and $C_2$ are harder to adapt to the driving system, which is also implied by the smaller values of information flow and information susceptibility. For driven $C_3$ or $C_4$, the noisy dependence was never observed and the values converged faster.

There are six cases in which $\epsilon_{max}$ is marked by bifurcations in the driven subsystem. In all these cases except $R_1\rightarrow C_4$, there is a discrete jump in the information flow at the point of system change.

In summary, we have shown that the coupling parameter of most systems is uniquely determined by the information flow in the limit of weak coupling. In measured time series, the models of the measured systems are usually unknown and surrogate data are often needed to determine the direction of coupling. In coupled Duffing systems, we can determine both the direction and the magnitude of the coupling in the limit of weak coupling. This is possible because the model of the systems is known, or rather known not to change. Only the coupling parameter changes, while the other system parameters are kept constant. If this was known to be the case in a realistic system, inferring the direction as well as the magnitude of coupling might be feasible solely by calculating the information flow.

\section{Conclusion}

One of the challenges in studying complex systems is to infer the coupling between individual systems from their time series. We studied the behavior of information-based coupling detection methods on coupled Duffing oscillator systems. The information flow successfully determined the direction of coupling between chaotic or regular systems. This was always possible for small coupling, and in most cases also for large coupling.

To find the numerical properties, the convergence of \gls{cmi} with the length of the time series and the dependence on different partitions were observed. The convergence was found to be faster than expected, and a finer partition to be generally beneficial as long as the individual bins are well populated.

While the absolute value of \gls{cmi} is considered of little value in experimental data, we can analyze it thoroughly in mathematical models. Thus, we studied the dependence of \gls{cmi} on the coupling parameter. We evaluated the usefulness of \gls{cmi} in solving the inverse problem of inferring the coupling parameter. By defining the information susceptibility, we found that in some cases the coupling parameter is uniquely determined by the information flow.
This is the case in the limit of strong coupling, and at least in some cases in the limit of weak coupling. Although this knowledge cannot be applied to arbitrary experimental data, it might be possible if accurate models of the systems were known.

The inverse problem proved to be complex because bifurcations can occur when systems are coupled. The exact classification of systems for which it is solvable remains to be seen.

\begin{acknowledgments}
The authors MB and PB acknowledge the financial support from the Slovenian Research Agency (research program No.\ P2--0001). MH acknowledges the financial support from the Slovenian Research Agency (research program No.\ P1--0402 and research projects No.\ J1--1698).
\end{acknowledgments}

\appendix

\section{Convergence of \gls{cmi}}
\label{appendix convergence}

The expected convergence of histograms with time $p(t) = p_0 + At^{-C}$ is $C=1/2$ for chaotic and $C=1$ for regular systems. In \figurename~\ref{convergence} it was found that the convergence of the information flow is $C=1$ for all systems. To find out why this is the case, we tested the convergence of different variables for the system $C_2 \rightarrow C_1$. For this test, we chose $\epsilon_1 = 0$, since we expect \gls{cmi} to converge close to $0$.

\begin{figure}
\includegraphics[width = 0.95\linewidth]{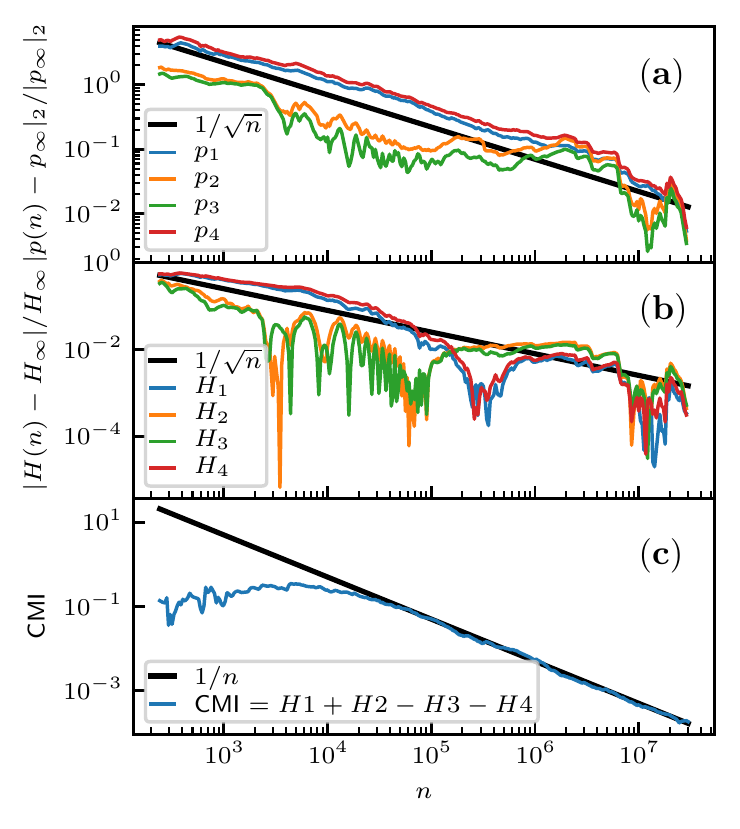}
\caption{The convergence of probabilities (a), entropies (b) and conditional mutual information (time-lagging $C_2$ with $\tau=1$) (c) with the length of the time series $n$ for the system $C_2 \rightarrow C_1$ with $\epsilon_1 = 0$. The black lines are given to indicate the type of convergence.}
\label{convergence2}
\end{figure}

We looked at the convergence of the conditional mutual information \eqref{CMI2} towards $0$, the four entropies involved in the calculation of \gls{cmi} \eqref{cminum} and the probability distributions used to calculate these entropies. In \figurename~\ref{convergence2}, the probability distributions are denoted as $p_1$ to $p_4$ and the entropies as $H_1$ to $H_4$, according to the order in \eqref{cminum}. The entropy $H_3$ has two arguments, the entropies $H_1$ and $H_2$  have four arguments, and the entropy $H_4$ has six arguments. The same holds for the corresponding distributions.

As expected, the empirical probability distributions $p_i$ converge with $C = 1/2$. This was verified by treating the distributions as vectors and observing their convergence with increasing length of the series $n$ in 2-norm $|\cdot |_2$. The converged values are denoted by $p_{\infty}$ and $H_{\infty}$. The entropies $H_i$ also converge slower than with $C=1$. However, taking the sum and difference of the entropies defining the information flow \gls{cmi}, the convergence is evidently faster than the constituting terms with $C = 1$.

\section{Systems with bifurcations}
\label{appendix systems}

Some systems exhibit bifurcations, i.e., sudden qualitative changes in the system behavior caused by a small change in the coupling parameter $\epsilon_1$. In such cases, observing the phase space and conditional Lyapunov exponents helps in understanding the dependence of the information flow and the solvability of the inverse problem.

\subsection{System $R_2 \rightarrow C_1$}
\label{appendixR2vC1}

\begin{figure}
\includegraphics[width = 1.05\linewidth]{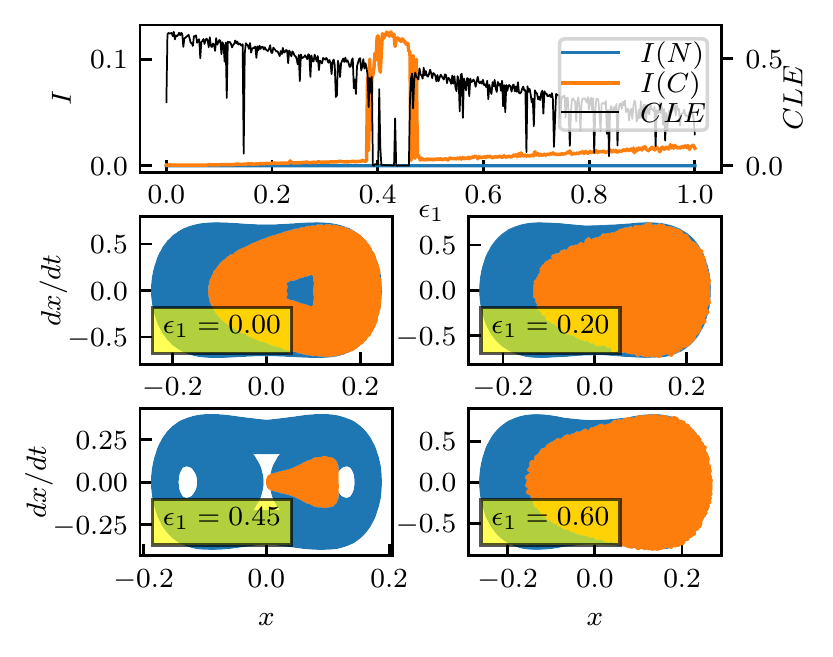}
\caption{The information flow and CLE dependences on the coupling parameter and the corresponding phase space portraits of the driven subsystem for $R_2 \rightarrow C_1$. The orange dots represent a stroboscopic map at $\cos(\omega_1 t) = 1$.}
\label{R2vC1}
\end{figure}

The information flow dependence, the CLE dependence, and the phase space portraits for the system $R_2 \rightarrow C_1$ are shown in \figurename~\ref{R2vC1}. The inverse problem is solvable for $\epsilon_1 \leq 0.38$, where the dynamics of the system change. The change can also be seen from smaller values of the CLE.

\begin{figure}[h!]
\includegraphics[width = 1.03\linewidth]{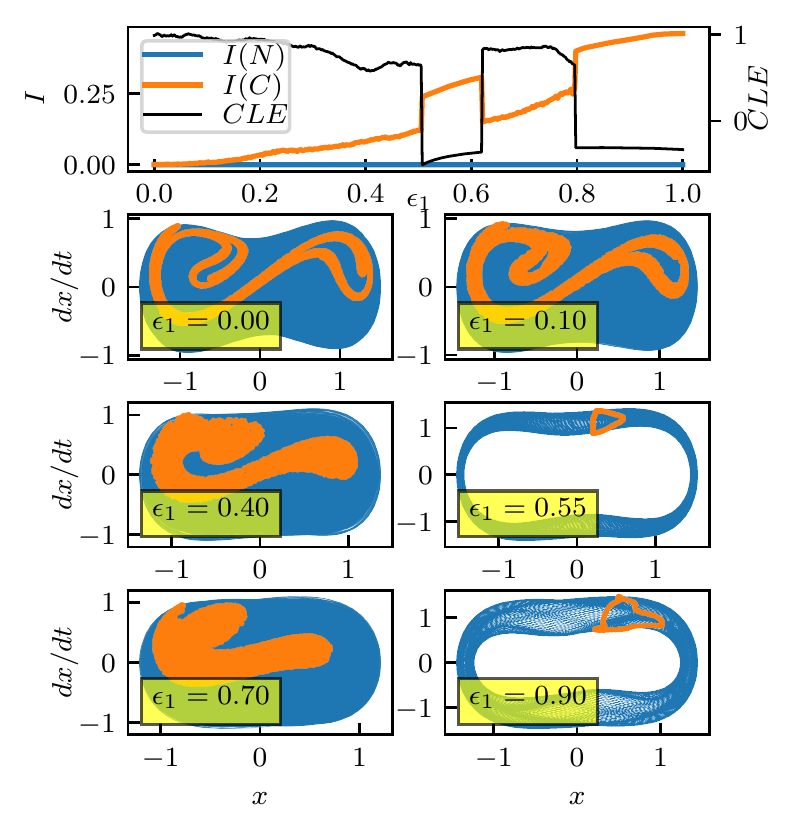}
\caption{The information flow and CLE dependences on the coupling parameter and the corresponding phase space portraits of the driven subsystem for $R_2 \rightarrow C_4$. The orange dots represent a stroboscopic map at $\cos(\omega_1 t) = 1$.}
\label{R2vC4}
\end{figure}

\subsection{System $R_2 \rightarrow C_4$}
\label{appendixR2vC4}

The information flow dependence, the CLE dependence, and the phase space portraits for the system $R_2 \rightarrow C_4$ are shown in \figurename~\ref{R2vC4}. In the regions $0.51 <\epsilon_1 < 0.61$ and $\epsilon_1 < 0.8$, the driven subsystem exhibits regular dynamics and the information flow takes higher values, while the CLE takes negative values. The inverse problem is solvable until the first appearance of regular dynamics at $\epsilon_{max} = 0.51$.

\subsection{System $C_2 \rightarrow C_3$}
\label{appendixC2vC3}

\begin{figure}
\includegraphics[width = 1.08\linewidth]{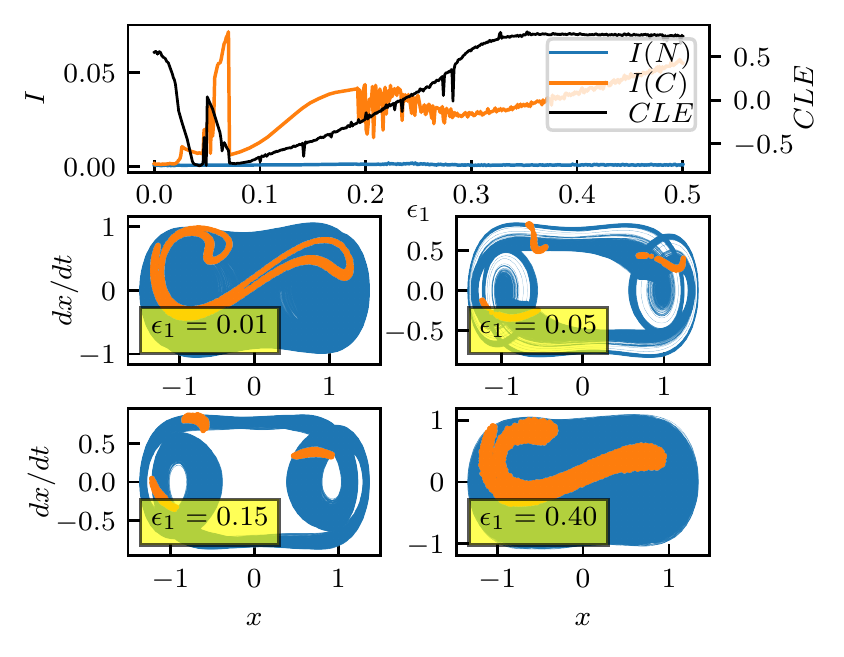}
\caption{The information flow and CLE dependences on the coupling parameter and the corresponding phase space portraits of the driven subsystem for $C_2 \rightarrow C_3$. The orange dots represent a stroboscopic map at $\cos(\omega_1 t) = 1$.}
\label{C2vC3}
\end{figure}

The information flow dependence, the CLE dependence, and the phase space portraits for the system $C_2 \rightarrow C_3$ are shown in \figurename~\ref{C2vC3}. The system dynamics change several times, the first time at $\epsilon_{max} = 0.021$. This is seen in an intense behaviour of the information flow and the CLE values.

\subsection{System $R_1 \rightarrow C_4$}
\label{appendixR1vC4}

\begin{figure}
\includegraphics[width = 1.06\linewidth]{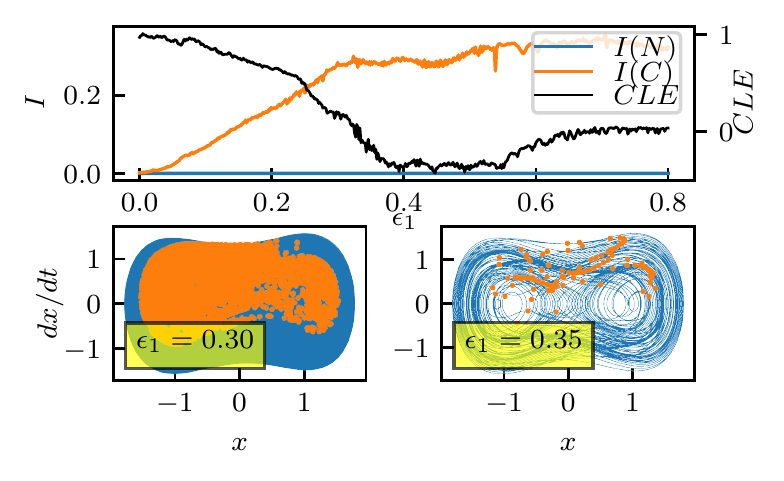}
\caption{The information flow and CLE dependences on the coupling parameter and the corresponding phase space portraits of the driven subsystem for $R_1 \rightarrow C_4$. The orange dots represent a stroboscopic map at $\cos(\omega_1 t) = 1$.}
\label{R1vC4}
\end{figure}

The information flow dependence, the CLE dependence, and the phase space portraits for the system $R_1 \rightarrow C_4$ are shown in \figurename~\ref{R1vC4}.
The inverse problem is solvable up to $\epsilon_{max} = 0.32$, where the dynamics of the driven subsystem become regular, which is reflected in the negative values of the CLE. The driven subsystem becomes periodic with a period of exactly $100$, i.e. $t_{period} = 100 \cdot 2\pi/\omega_1$.

\subsection{System $R_2 \rightarrow C_3$}
\label{appendixR2vC3}

\begin{figure}
\includegraphics[width = 1.05\linewidth]{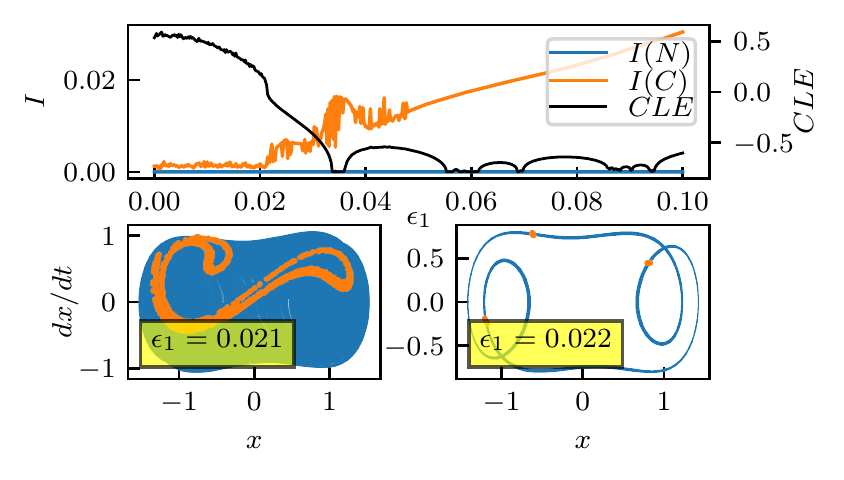} 
\caption{The information flow and CLE dependences on the coupling parameter and the corresponding phase space portraits of the driven subsystem for $R_2 \rightarrow C_3$. The orange dots represent a stroboscopic map at $\cos(\omega_1 t) = 1$.}
\label{R2vC3}
\end{figure}

The information flow dependence, the CLE dependence, and the phase space portraits for the system $R_2 \rightarrow C_3$ are shown in \figurename~\ref{R2vC3}. The dynamics of the driven subsystem become regular at $\epsilon_{max} = 0.022$, which is reflected in the negative values of the CLE.

\begin{figure}
\includegraphics[width = 1.07\linewidth]{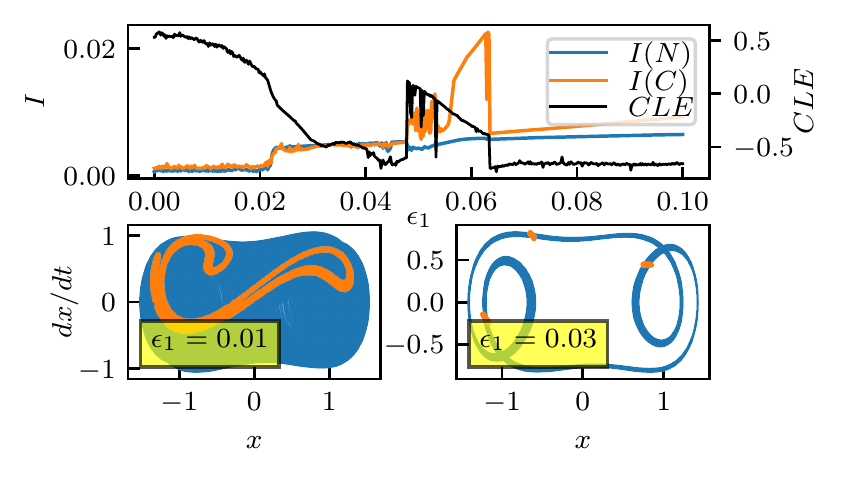} 
\caption{The information flow and CLE dependences on the coupling parameter and the corresponding phase space portraits of the driven subsystem for $C_1 \rightarrow C_3$. The orange dots represent a stroboscopic map at $\cos(\omega_1 t) = 1$.}
\label{C1vC3}
\end{figure}

\subsection{System $C_1 \rightarrow C_3$}
\label{appendixC1vC3}

The information flow dependence, the CLE dependence, and the phase space portraits for the system $C_1 \rightarrow C_3$ are shown in \figurename~\ref{C1vC3}. The dynamics of the driven subsystem become regular at $\epsilon_{max} = 0.023$, which is reflected in the negative values of the CLE.

\section*{Data availability}

The data that support the analysis of this article have been generated by the authors and can be fully reproduced from the repository \url{https://repo.ijs.si/mbresar/duffing-cmi} \cite{repo}.

\bibliography{detcoupling}

\end{document}